# A Non-Isolated High Step-Up Interleaved DC-DC Converter with Diode-Capacitor Multiplier Cells and Dual Coupled Inductors

Mamdouh L. Alghaythi[1], Robert M. O'Connell[1], Naz E. Islam[1], and Josep M. Guerrero[2]

[1]Department of Electrical Engineering and Computer Science, University of Missouri, Columbia, USA
[2]Center for Research on Microgrids (CROM), Department of Energy Technology, Aalborg University,
9220 Aalborg East, Denmark
mlagzd@mail.missouri.edu[1], OConnellR@missouri.edu[1], islamn@missouri.edu[1], and joz@et.aau.dk[2]

*Abstract*—In this paper, a non-isolated high step-up dc-dc converter is presented. The proposed converter is composed of an interleaved structure and diode-capacitor multiplier cells for interfacing low-voltage renewable energy sources to high-voltage distribution buses. The aforementioned topology can provide a very high voltage gain due to employing the coupled inductors and the diode-capacitor cells. The coupled inductors are connected to the diode-capacitor multiplier cells to achieve the interleaved energy storage in the output side. Furthermore, the proposed topology provides continuous input current with low voltage stress on the power devices. The reverse recovery problem of the diodes is reduced. This topology can be operated at a reduced duty cycle by adjusting the turn ratio of the coupled inductors. Moreover, the performance comparison between the proposed topology and other converters are introduced. The design considerations operation principle, steady-state analysis, simulation results, and experimental verifications are presented. Therefore, a 500-W hardware prototype with an input voltage of 30-V and an output voltage of 1000-V is built to verify the performance and the theoretical analysis.

## I. INTRODUCTION

THE environmental concerns, such as global warming, climate change and air pollution is becoming a critical public apprehension. Nowadays, renewable energy sources, such as photovoltaics (PVs) and fuel cell are promising alternative energy sources and extensively utilized in many applications. However, the output voltage of PV panels is low in the end stage and depends on the weather conditions, such as the change in temperature and solar irradiance. Power electronic converters play a vital role in power conversion of the distributed generation and the grid [1]. Thus, due to the aforementioned shortcoming, employing high step-up dc-dc converters plays a substantial role in increasing and regulating the low voltage of PV panels to high voltage [2],[3].

The conventional boost converter can be employed to boost the output voltage but that would require using extreme duty cycles and turns ratios. However, employing extreme duty cycles would eventually cause the voltage stresses on passive components and the conduction losses, and the voltage stress of the semiconductors is as excessive as the output voltage. The cascaded boost converter can be considered as a proper applicant for these high step-up gain converters. Nevertheless, applying cascaded boost converter will result in increasing the size of the converter and the cost and cause the electromagnetic interference problems. Moreover, it would reduce the reliability and efficiency and of the converter, and the output diode may experience a high voltage stress on the switches [4].

Many high voltage gain converters have been proposed with different voltage boost techniques to overcome the aforementioned shortcomings and enhance the voltage gain. In [5], a novel coupled inductor-based high step-up dc–dc converter has been proposed. In additions, the power density is increased due to high switching frequency, the voltage spike of the switch is clamped during the turn-off condition due to employing the clamped circuit, and zero voltage switching (ZVS) is given due the exitance of the energy in the leakage inductance. A high step-up interleaved dc-dc converter with asymmetric voltage multiplier and coupled inductors has been introduced in [6]. The aforementioned converter can be utilized to reduce the switching losses due to using the zero current switching (ZCS), lift the voltage gain and minimize the input current ripple, which would increase the lifetime of the input power. In [7], a group of cascaded high gain dc-dc converters with clamped circuits has been proposed to provide an ultra high voltage conversion with minimizing the passive components stresses. Several high step-up dc-dc topologies with ZVS turn-on capability are presented to achieve ultrahigh gain and mitigate and power devices losses in [8]-[13]. The losses in those topologies have been reduced due to the low resistance *r-ds,* and the reverse recovery problems have been alleviated due to the active clamp circuits [14]-[19]. Many high voltage gain dc-dc converters with coupled inductor and voltage multiplier techniques have been introduced [20]-[25]. Employing transformers can increase the output voltage level but cause result in decreasing the power density and cause the leakage inductance [26]-[31].This paper introduces the operational principle of the proposed converter in section II, operating modes in part A of section II, voltage gain derivation in part B of section II, stress on power devices and performance comparison in section III, voltage stress and simulation results in part A of section III, experimental verifications in part B of section III, and conclusion in section IV.

## II. OPERATIONAL PRINCIPLE OF THE PROPOSED CONVERTER

This section primarily presents the operational principle of the proposed converter. In addition, the aforementioned converter comprises the interleaved boost converter with coupled inductors in the primary side, and the voltage multiplier cells with the secondary windings in the secondary side is as depicted in Fig. 1. Additionally, the current ripple and the losses can be reduced, and the voltage gain in the proposed structure can be extend further. The interleaved boost converter is mainly consisted of two coupled inductors, two magnetizing inductors $L_{m1}$ and $L_{m2}$, two leakage inductances $L_{k1}$ and $L_{k2}$ and two switches connected to the source. The voltage multiplier cells are connected with the secondary windings of the coupled inductors. The switching losses are reduced due to employing the zero current switching (ZCS). The input current ripple decreases as the phases of the interleaved converter increase. The phase shift between the switches is 180 degree, and the switching signals waveforms are depicted in Fig. 2.of the. As a result, in order to simplify the circuit, The following assumptions are considered

(1) The power devices are ideal, but the leakage inductances of the coupled inductors are modeled ideally with a magnetizing inductor a leakage inductor and an ideal transformer with a turns ratio of $N = \frac{N_{1b}}{N_{1a}} = \frac{N_{2b}}{N_{2a}}$ (1)

(2) All Capacitors are large enough to keep the voltage of the capacitors constant and make their voltages ripples free.

(3) The current ripple in the inductors is negligible, and the inductors are large enough. The analysis of the proposed high voltage gain dc-dc converter is under the continuous conduction mode.

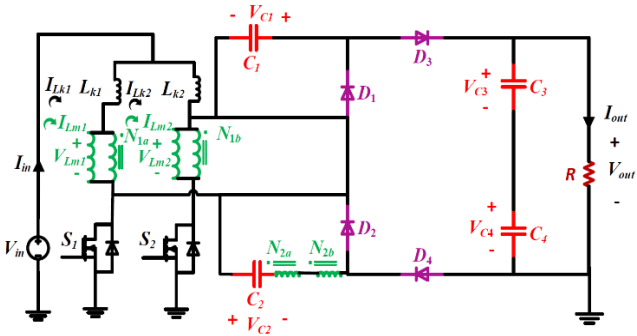

Fig. 1. The proposed high voltage gain dc-dc converter.

### A. Operating Modes

The operation modes of the proposed converter are introduced as follows.

*Mode 1*: $S_1$ and $S_2$ are ON as depicted in Fig. 3. $L_{m1}$ and $L_{m2}$ store the energy, and the magnetizing inductors $L_{m1}$ and $L_{m2}$, the leakage inductances $L_{k1}$ and $L_{k2}$ are charged by the source. The magnetizing inductor currents are linearly increased. The diodes of voltage multiplier (VM) do not conduct, and they are reverse biased. fixed. Finally, the load energy is supplied by $C_3$ and $C_4$.

*Mode 2*: In the second mode, $S_1$ is OFF and $S_2$ is ON as shown in Fig. 4. $D_2$ and $D_3$ do not conduct, and they are reverse biased. However, $D_1$ and $D_4$ are forward biased and conduct in this mode. The magnetizing inductance $L_{m1}$, the leakage inductance $L_{k1}$ and the source charge $C_2$ and $C_3$. $L_{m2}$ and $L_{k2}$ are both charged by the source. Thus, $L_{m2}$ stores the energy, and $L_{m1}$ release the energy.

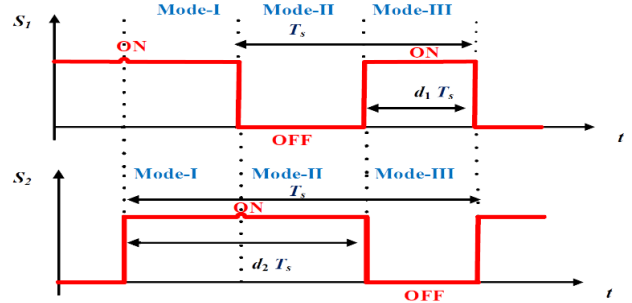

Fig. 2. Switching signals of the proposed topology.

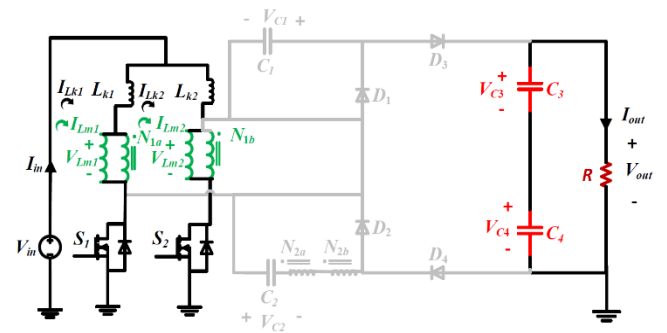

Fig. 3. Equivalent circuit of Mode 1.

*Mode 3*: $S_1$ is ON and $S_2$ is OFF. $D_1$, and $D_4$ are reverse biased. However, $D_2$ and $D_3$ are conducting, and they are forward biased. $L_{m2}$ and the input voltage charge $C_1$ and $C_4$. $L_{m1}$ stores the energy, and the magnetizing inductance $L_{m2}$, and the leakage inductance $L_{k2}$ transfers the energy.

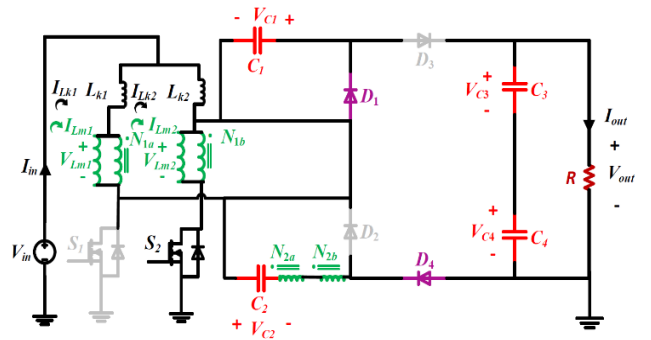

Fig. 4. Equivalent circuit of Mode 2.

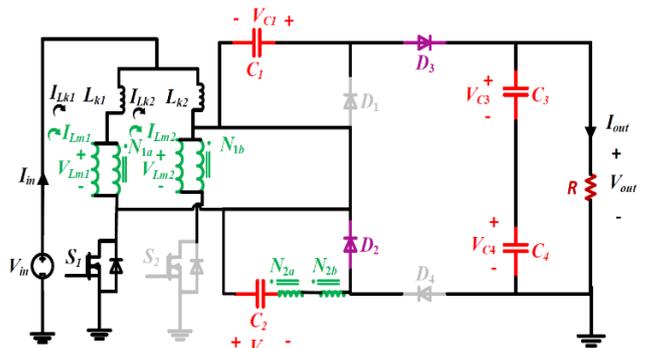

Fig. 5. Equivalent circuit of Mode 3.

## B. Voltage Gain Derivation

This section introduces the derivation of the voltage gain. For the second mode, the steady state equations can be expressed as

$$< V_{Lm1} >=< V_{Lm2} >= 0 \quad (2)$$
$$V_{in} = \frac{dI_{Lm1}}{dt} \quad (3)$$
$$V_{in} = \frac{dI_{Lm2}}{dt} \quad (4)$$

The capacitor voltages for $C_1$ and $C_2$ are equal, the magnetizing voltages are identical and can be derived as

$$V_{Lm1} = V_{Lm2} = V_{in} + V_{c1} - V_{c2} - V_{c3} - V_{c4} \quad (5)$$
$$V_{Lm1} = V_{Lm2} = V_{in} \quad (6)$$

The volt-second balance principle, and the number of turns ratio is 1.5

$$M = \frac{V_{out}}{V_{in}} = 2(N+3) \times \frac{1}{(1-D)} \quad (7)$$

For mode 3, the capacitor voltages can be written as

$$V_{C1} = V_{in} \times \frac{1}{(1-D)} \quad (8)$$
$$V_{C2} = V_{in} \times \frac{2}{(1-D)} \quad (9)$$
$$V_{C3} = V_{in} \times \frac{3}{(1-D)} \quad (10)$$
$$V_{C4} = V_{in} \times \frac{4}{(1-D)} \quad (11)$$

The voltage gain conversion can be written as

$$V_{out} = V_{in} \times \frac{9}{(1-D)} \quad (12)$$

## III. STRESS ON POWER DEVICES AND PERFORMANCE COMPARISON

### A. Voltage Stress and Simulation Results

This section discusses the voltage stress on components and comparison analysis. Hence, the voltage stress across the switches can be given as

$$V_{s1} = V_{s2} = V_{in} \times \frac{1}{(1-D)} \quad (13)$$

The magnetizing inductor currents and the output current can be derived as

$$I_{Lm1(avg)} = I_{Lm2(avg)} = \frac{(N+2) \, I_{out}}{(1-D)} \quad (14)$$
$$I_{out} = \frac{V_{out}}{R} \quad (15)$$

It can be observed from Fig. 6 and Fig. 7 that the voltage stress on the switches is reduced to 111.11 V, the magnetizing inductor currents are 20.20 A, the input voltage is 30 V and the output current is 1.21 A. Additionally, the voltage stress on diodes is reduced to 222.22 V, and the diode current are identical to the output current as depicted in Fig. 8. The voltage stress on diodes can be expressed as

$$V_{Dmax} = V_{D1} = V_{in} \times \frac{2}{(1-D)} \quad (16)$$
$$V_{D2} = V_{D3} = V_{D4} = V_{in} \times \frac{2}{(1-D)} \quad (17)$$

The input current is the total of the two magnetizing inductor currents and can be calculated as

$$I_{in} = \frac{2 \, (N+2) \, I_{out}}{(1-D)} \quad (18)$$

The input current is 40.40 A, the input voltage is 30 V, the voltage stress on $C_1$ is 111.11 V and the output voltage is lifted to 1000 V as illustrated in Fig. 9.
The voltage stresses on $C_2$, $C_3$ and $C_4$ are 222.22 V, 333.33 V, 444.44 V as illustrated in Fig. 10.

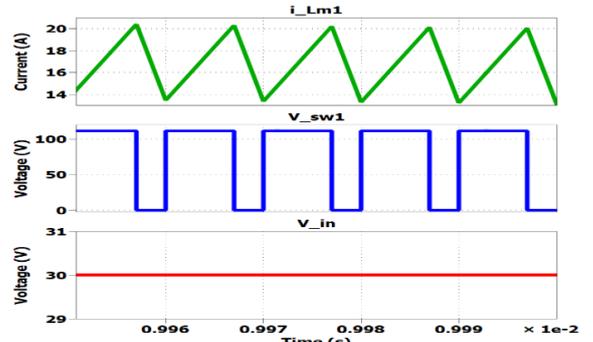

Fig. 6. The magnetizing inductor current, voltage stress on the first switch and the input voltage.

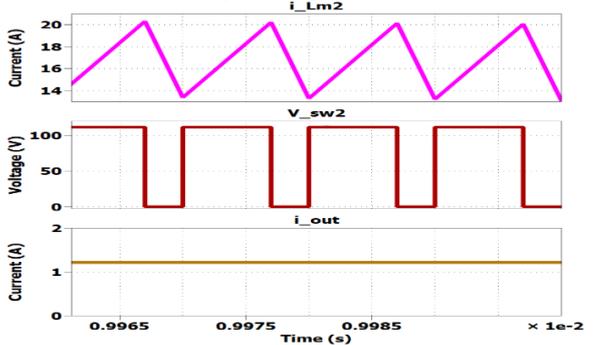

Fig. 7. The magnetizing inductor current, voltage stress on the second switch the output current.

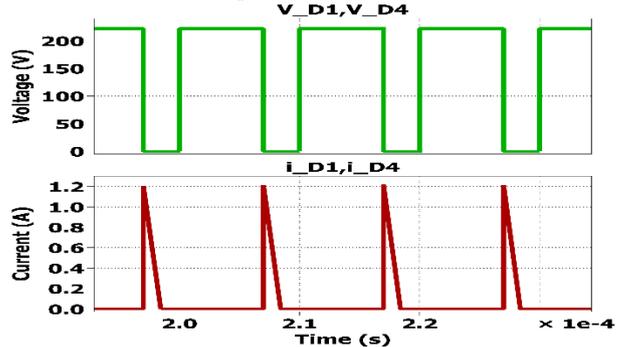

Fig. 8. Voltage and current stress on diodes.

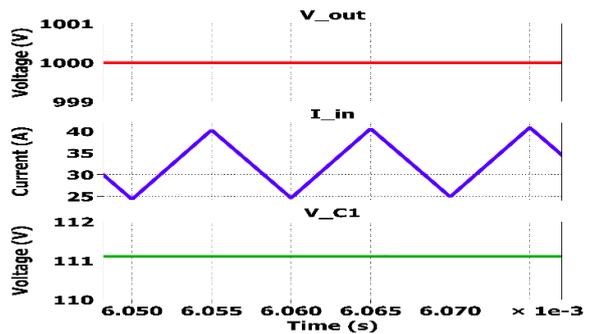

Fig. 9. The output voltage, the input current, and the voltage stress on the first capacitor.

TABLE I illustrates the performance between the proposed topology and other topologies in terms of voltage gain and voltage stress on switches, voltage stress on diodes, sharing ground connection, the input current and the total number of passive components and inductor cores. It can be noted that the voltage stress on switches and the total number of switches and capacitors in [5] and [6] are identical to the proposed converter. The total number of the inductor cores in [6] ground side for being floated and not shared is identical to the proposed topology. The voltage stress on the switches and diodes in [10] is high and could cause high conduction losses. The total number of power devices in [11] is equal to the proposed topology. TABLE II shows the selected components for the simulation part.

TABLE I
PERFORMANCE COMPARISON BETWEEN THE PROPOSED CONVERTER AND OTHER TOPOLOGIES

| Topologies | Voltage gain | Voltage stress on switches | Voltage stress of diodes | No. of components | | | No. of inductor cores | | Shared grounded | Input Current |
|---|---|---|---|---|---|---|---|---|---|---|
| | | | | S | D | C | Single | Coupled | | |
| Ref. [5] | $\dfrac{2N-1}{(N-1)(1-D)}$ | $\dfrac{V_{in}}{(1-D)}$ | $\dfrac{NV_o}{2N-1}$ | 2 | 2 | 4 | 1 | 1 | Yes | Continuous |
| Ref. [6] | $\dfrac{3N+1}{1-D}$ | $\dfrac{V_o}{3N+1}$ | $\dfrac{2NV_o}{3N+1}$ | 2 | 5 | 4 | 0 | 2 | No | Continuous |
| Ref. [10] | $\dfrac{1}{1-2D}$ | $V_o$ | $V_o$ | 1 | 2 | 3 | 2 | 0 | Yes | Discontinuous |
| Ref. [11] | $\dfrac{2(N+1)}{1-D}$ | $\dfrac{V_o}{2(N+1)}$ | $\dfrac{NV_o}{N+1}$ | 2 | 4 | 4 | 0 | 2 | Yes | Continuous |
| Proposed converter | $\dfrac{2N+2}{1-D}$ | $\dfrac{V_{in}}{(1-D)}$ | $\dfrac{2V_{in}}{(1-D)}$ | 2 | 4 | 4 | 0 | 2 | No | Continuous |

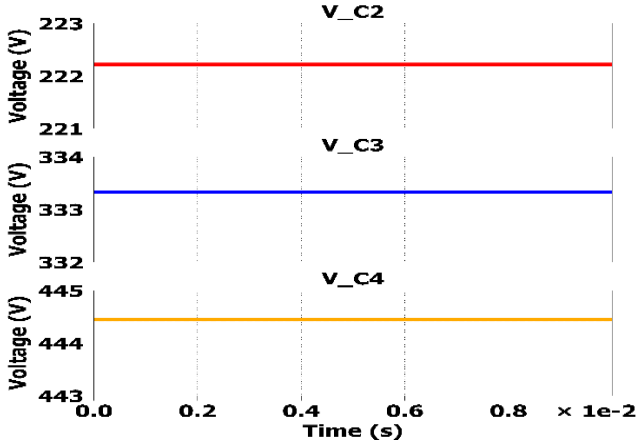

Fig. 10. The voltage stress on the capacitors.

TABLE II
COMPONENT PARAMETERS

| Parameter | Value |
|---|---|
| Input Voltage | 30 V |
| Output Voltage | 1000 V |
| Duty Cycle | 0.73 |
| Load Resistance | 825 Ω |
| Magnetizing Inductors | 94 µH |
| Frequency ($f_{sw}$) | 118 KHZ |
| Capacitors | 10µF |

*B. Experimental Verifications*

This section provides the experimental validation for the design and the theory. Additionally, a 500 W hardware prototype has been built to verify the simulation results and the design. The hardware prototype specifications are described in TABLE III. The topology is rated at 500 W with input voltage of 30 V and output voltage of 1000 V. The effective performance DPG20C300PN diodes are selected to achieve the voltage blocking on diodes, and the active switches are attained by employing MOSFETs IPA075N15N3GXKSA1. In addition, the capacitors are carefully chosen by utilizing MKP338 1 X1 to execute the existence high voltage blocking. The magnetizing inductor is done by using coupled inductors and magnetic core of EDT31 ferrite core, B66397 and N87 material. It can be observed that the parasitic resistance of the components could cause the conduction losses and can be written as

$$P_{sw\,(cond\,loss)} = (I_{sw(rms)} \times I_{sw(rms)}) \times R_{DS(on)} \quad (19)$$

The equivalent series resistance (ESR) of the capacitor may cause the capacitors losses and can be derived as

$$P_{C\,(loss)} = (I_{C(rms)} \times I_{C(rms)}) \times ESR \quad (20)$$

The diodes losses include the forward voltage and the average diode currents and can be given as

$$P_{D\,(loss)} = I_{D(avg)} \times V_F \quad (21)$$

TABLE III
HARDWARE PROTOTYPE SPECIFICATIONS

| Components | Description |
|---|---|
| MOSFETs ($S_1, S_2$) | IPA075N15N3GXKSA ($R_{ds(on)}$ =7.5mΩ, 150V, 43 A) |
| Diodes ($D_1$–$D_4$) | DPG20C300PN (300 V, 10A, 35 ns) |
| Capacitors ($C_1 - C_4$) | MKP338 1 X1 (1 µF, 1000V) |
| Coupled Inductor ($L_{m1}, L_{m2}$) | EDT31 ferrite core, B66397 (94 µH, N87 material) |

The experimental results of the voltage stress on the switches, input current, and the output voltage as depicted in Fig.11. The voltage stress on the switches is 111.11 V, the input current is 40.40 A and the output voltage is 1000V. The peak efficiency of the proposed converter is 96.9% at the output power of 500 W in different loads. Moreover, it can be observed that the maximum efficiency occurs at 500 W as shown in Fig. 12.

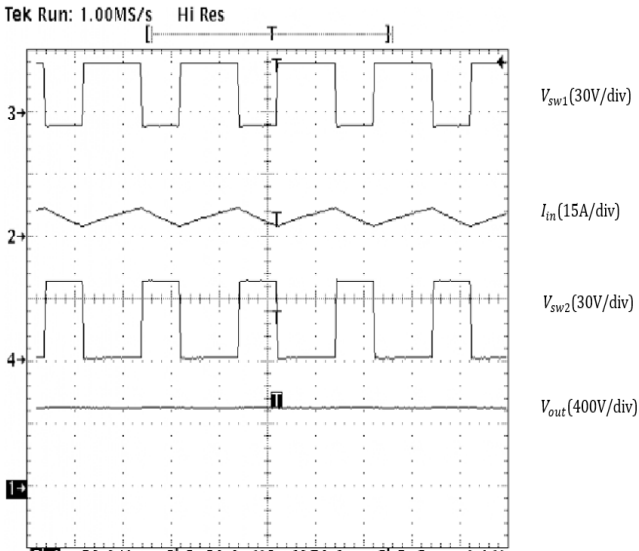

Fig. 11. Experimental waveforms of the voltage stress on switches, input current, and the output voltage.

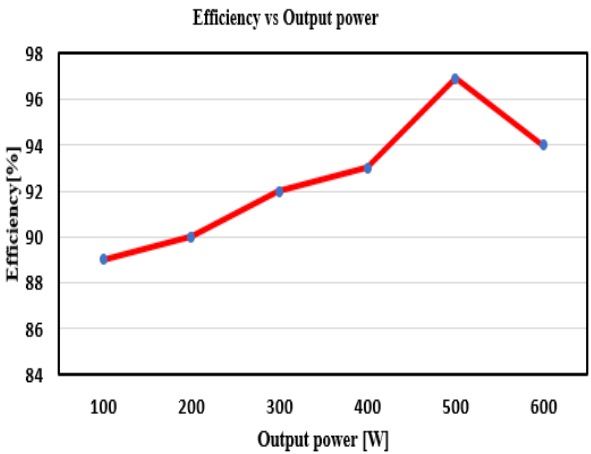

Fig. 12. Efficiency [%] vs. the output power [W].

The power loss breakdown of the proposed converter is depicted in Fig. 13, and it can be noted that the coupled inductor loss are the dominant power losses, the diode loss, the capacitor loss and the switch loss. Thus, more coppers are highly recommended to enhance the efficiency of the proposed converter.

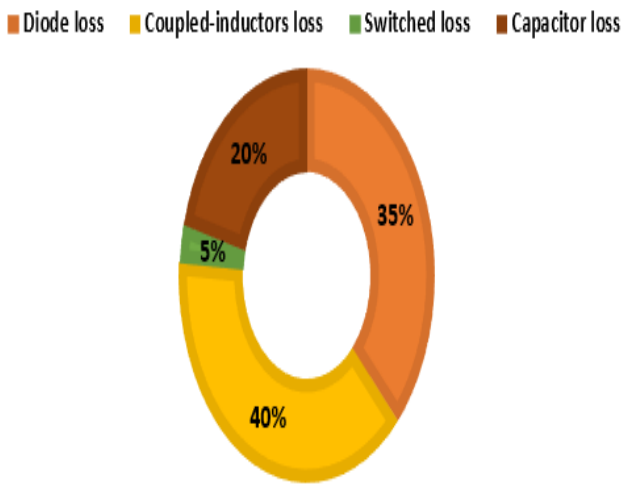

Fig. 13. The power loss breakdown.

## IV. Conclusion

A non-isolated high step-up dc-dc converter with diode-capacitor multiplier cells and dual coupled inductors has been introduced in this paper. Moreover, the input current of the proposed converter was continuous with small ripples due to the interleaved structure. The aforementioned topology has accomplished an ultra-voltage gain and power ability due to the coupled inductors and the voltage multiplier cell without an extreme duty cycle or a high turn ratio. The voltage gain, voltage stress on switches, voltage stress on diodes, sharing ground connection, the input current and the total number of passive components were compared between the proposed converter and other recent high step-up dc-dc converters. In addition, the voltage stress on power devices, the conduction losses and the cost were decreased. The reverse recovery problem of the diodes was alleviated, and the leakage energy was recycled. The theoretical analysis, operational principles, simulation results and experimental outcomes were designed. A hardware prototype was constructed to verify the design and the theory with improving the efficiency.